%% file: main.tex
\documentclass[nonacm,sigplan,screen]{acmart}

\AtBeginDocument{%
  }

\usepackage{graphicx} 
\usepackage{todonotes}
\usepackage{paralist}

\definecolor[named]{lipicsLightGray}{rgb}{0.85,0.85,0.86}

\usepackage{listings}
\usepackage{cuted} 
\usepackage{enumitem}
\usepackage{url}

\lstnewenvironment{dafny}[1][]{\lstset{language=dafny,basicstyle=\ttfamily\footnotesize, numbers=left, stepnumber=1, numberstyle=\tiny, numbersep=1pt, #1}}{}
\lstdefinelanguage{dafny}{
  sensitive=true,
  alsoletter={\%},
  comment=[l]{//},
  morecomment=[s]{/*}{*/},
  commentstyle=\color{gray},
  string=[b]{"},
  morekeywords=[1]{abstract,allocated,as,break,case,%
    continue,else,exists,expect,false,for,%
    forall,fresh,if,in,is,label,%
    match,modify,nameonly,new,null,object,object?,old,%
    print,refines,return,reveal,%
    static,then,this,true,unchanged,while,%
    yield,yields,by,calc,extern,axiom},%
        keywordstyle=[4]\color{blue},
        keywordstyle=[3]\color{blue},
        keywordstyle=[2]\color{blue},
       keywordstyle=[1]\color{blue},
  morekeywords=[2]{nat,int,set,multiset,iset,map,imap,bool,char,ORDINAL,real,%
    seq,string},
  morekeywords=[3]{import,opened,export,reveals,provides,include},
  morekeywords=[4]{assert,assume,decreases,ensures,invariant,modifies,%
    reads,requires,witness},
  morekeywords=[4]{class,codatatype,const,var,constructor,datatype,function,%
    iterator,lemma,method,newtype,predicate,trait,type,module,least,greatest,%
    ghost,extends,opaque,returns,twostate},
    }

\lstdefinestyle{inlinedafny}{
  language=dafny,
  basicstyle=\ttfamily
}

\def\inl{\lstinline[style=inlinedafny]}

\newif\ifoutline
\outlinefalse

\newcommand{\contents}[1]{\ifoutline{\footnotesize\color{blue}
    \begin{itemize}
    #1
    \end{itemize}
  \par}\bigskip\fi}

\setcopyright{none}
\renewcommand\footnotetextcopyrightpermission[1]{}

\begin{document}

\title{Verification of E-Voting Algorithms in Dafny}

\author[Robert Büttner, Fabian Franz Dießl, Patrick Janoschek, Ivana Kostadinovic, Henrik Oback, Kilian Voß, Franziska Alber, Roland Herrmann, Sibylle Möhle, Philipp Rümmer]{Robert Büttner, Fabian Franz Dießl, Patrick Janoschek, Ivana Kostadinovic, Henrik Oback,\\ Kilian Voß, Franziska Alber, Roland Herrmann, Sibylle Möhle, Philipp Rümmer}
\affiliation{%
  \institution{University of Regensburg}
  \city{Regensburg}
  \country{Germany}
}



\begin{abstract}
  Electronic voting procedures are implementations of electoral systems, making it possible to conduct polls or elections with the help of computers. This paper reports on the development of an open-source library of electronic voting procedures, which currently covers Score Voting, Instant-Runoff Voting, Borda Count, and Single Transferable Vote. The four procedures, of which two are discussed in detail, have been implemented in Dafny, formally verifying the consistency with functional specifications and key correctness properties. Using code extraction from the Dafny implementation, the library has been used to set up a voting web service.



\end{abstract}
    
\maketitle


\section{Introduction}

\contents{
\item Motivation, what are voting algorithms, why are they important
\item There are few/hardly any verified implementations of voting algorithms
\item what makes the verification of voting algorithms challenging?
\item Which algorithms we consider and what we did, Dafny~\cite{DBLP:conf/lpar/Leino10}
\item Contributions
}

Elections and voting are fundamental mechanisms for collective decision making in modern societies. While elections are traditionally carried out by analog means such as casting paper ballots and manual counting, they are today increasingly implemented through electronic voting (\emph{e-voting}), making it possible for voters to cast their votes on a computer, or online, and automating the process of counting the votes. The correctness of e-voting procedures is therefore crucial, as bugs in the algorithms or implementations could lead to votes not being represented correctly in the results of an election, compromising trust in the elections themselves~\cite{DBLP:journals/cacm/FrancoPSV04,DBLP:conf/voteid/OttoboniS19,socialElections}.

There is a growing body of research on formally verifying properties of the \emph{electoral systems} realized by e-voting procedures. However, relatively little work has focused on verifying the actual \emph{implementations} of e-voting systems. In this paper, we report on a project developing a Dafny~\cite{DBLP:conf/lpar/Leino10} library of verified e-voting procedures, which currently includes the single-winner methods \emph{Score Voting}~\cite{Smith2000}, \emph{Instant-Runoff Voting}~\cite{Tideman06}, and \emph{Borda Count}~\cite{Emerson13}, as well as the proportional \emph{Single Transferable Vote}~\cite{Tideman95} method. The four algorithms cover a broad spectrum of electoral systems, providing a representative basis for exploring the benefits and challenges of applying formal verification techniques to e-voting.

\paragraph{Contributions}
We present an open-source library implementing four e-voting algorithms, of which two (Instant-Runoff Voting and Single Transferable Vote) are discussed in detail in this paper. To the best of our knowledge, the library provides the first Dafny programs for e-voting.
The correctness of all four implementations was shown against a declarative functional description of the respective electoral scheme. In addition, we formally verified several end-to-end properties of the voting procedures.
Extracting executable programs from the verified Dafny methods, the implementations are used to set up an e-voting web service.\footnote{A collection of both projects with infrastructure as well as aggregated Dafny sources can be found here: \url{https://github.com/uuverifiers/e-voting}.}

\section{Related Work}

There exists a wealth of work concerned with different aspects of e-voting algorithms.
In social choice theory, the focus is on general properties of electoral systems rather than on verifying actual implementations of voting procedures.
Rossetta~\cite{Rossetta24} verified anonymity, monotonicity, and concordance using Isabelle/HOL~\cite{DBLP:books/sp/NipkowPW02}, and Holliday et al.~\cite{DBLP:conf/lori/HollidayNP21} provided a formalization in Lean~\cite{DBLP:conf/cade/MouraKADR15} for verifying general properties as well as properties inherent to specific voting systems, such as Condorcet consistency.
In contrast to the work presented here, no executable prototype is available.

A further focus is on security-related aspects, e.g., privacy, verifiability, and coercion-resistance.
A formal verification of the Estonian e-voting protocol using the Tamarin theorem prover~\cite{DBLP:conf/cav/MeierSCB13} has been presented by Baloglu et al.~\cite{DBLP:conf/asiaccs/BalogluBM024}.
Similarly, the Norwegian e-voting algorithm has been formalized by Cortier and Wiedling~\cite{DBLP:conf/post/CortierW12} in the applied-pi calculus~\cite{DBLP:conf/popl/AbadiF01}.
Campanelli et al.~\cite{DBLP:conf/iciw/CampanelliFMPV08} have implemented a mobile e-voting application and formally verified it using the partial model checker PaMoChSA~\cite{PaMoChSA}, focusing on the mobility property.

Meghzili et al.~\cite{DBLP:journals/ijsinnov/MeghziliCEAB22} have adopted blockchain technology in their implementation of an e-voting system.
The system was formalized by means of a hierarchical colored Petri net~\cite{Jensen2009}.
Blockchain technology may increase the robustness of the voting process and has been adopted in several countries.
An overview can be found in the survey by Vladucu et al.~\cite{DBLP:journals/access/VladucuDMR23}.

\section{Electoral Systems and Algorithms}

\contents{
\item E-voting and electoral systems in general
\item Properties one would like to establish about voting procedures
\item High-level description of the four algorithms we consider
}

Electoral systems provide formal rules for aggregating individual preferences into a collective outcome, and different systems may yield very different results depending on their design~\cite{Nurmi87}. Electronic voting refers to the use of digital technologies to conduct elections, with the aim of making the process more efficient, accessible, and scalable while still preserving the essential properties of fairness, accuracy, and voter anonymity~\cite{Alvarez14}.

When designing or analyzing electoral systems, one is typically interested in properties such as determinism (the procedure produces a unique, well-defined result), correctness (the implementation conforms to the mathematical definition of the algorithm), fairness (voters’ preferences are taken into account according to the rules), and robustness (the system behaves predictably even in corner cases, such as ties). Formal verification offers a principled approach to establishing such properties at the level of executable code, thereby eliminating the risk of subtle implementation errors~\cite{Wimmer19}.


In what follows, we give a high-level overview of the four voting methods, highlighting their differences. 

\paragraph{Score Voting}
Each voter assigns a numerical score to each candidate within a fixed range. The scores are summed across all ballots, and the candidate with the highest total is declared the winner. This method allows voters to express both their order of preference and the intensity of their support, providing more nuanced information than simple rankings or approvals~\cite{Smith2000}.

\paragraph{Instant-Runoff Voting (IRV)}
Voters provide a ranking, which forms a strict and total order over a subset of candidates. If no candidate obtains a majority of first-choice votes, the candidates with the fewest of such votes are eliminated. Ballots that ranked these candidates first are transferred to the next available preference. This elimination and redistribution process continues until one candidate reaches a majority or all candidates are eliminated. IRV emphasizes majority support but can be sensitive to the order of eliminations~\cite{Tideman06}.

\paragraph{Borda Count}
Each voter submits a ranking (like the one used in IRV). The algorithm then assigns points to the selected candidates according to their position in this ranking: with $n$ candidates, the highest-ranked candidate receives $n$~points, the second $n-1$, and so on, down to 1 for the last position. The points are summarized on all ballots, and the candidate with the highest score wins. To handle ties, an optional variant of Baldwin’s method~\cite{Baldwin26, Tideman06} may be applied. On creating an election, a parameter determines the maximal number of tied placements (starting from the first place) to be resolved. The lowest-scoring candidates are then progressively eliminated from the rankings until a unique outcome is obtained. Borda Count takes full ranking into account and tends to favor broadly acceptable candidates~\cite{Emerson13}.

\paragraph{Single Transferable Vote (STV)}
This is a multi-winner system, where a specified number of seats gets filled with a subset of candidates. Voters provide a ranking as in IRV. A quota is computed to determine how many votes are required for an election. Candidates reaching the quota are elected, and if they receive more votes than necessary, the surplus is redistributed fractionally according to the next preferences, using Gregory’s method~\cite{Gregory80, Farrell11}. If no candidate meets the quota, the one with the fewest votes is eliminated, and their votes are transferred. This procedure continues until all seats are filled. STV provides proportional representation and thus ensures that minority groups have fair representation within the seats~\cite{Tideman95}.

\section{Voting Procedures in Dafny}

In the subsequent sections, a brief overview of the implementation and verification process is presented, followed by a more detailed examination of the Instant-Runoff Voting and Single Transferable Vote algorithms.
\subsection{Methodology and Design Decisions}
\label{sec:methodology}
\contents{
\item First define the general strategy we are following: (1) develop a functional model of an algorithm, (2) write an implementation, (3) prove consistency of the model and the implementation, (4) prove additional properties of the model or the implementation, (5) extract code
\item Development, CI, testing
}

The first step was the development of pseudo-code specifications that capture the essential functionality of each algorithm while remaining close enough to an executable form to later serve as the foundation for the Dafny implementation. In addition, several design decisions had to be made regarding the representation of candidates and votes in Dafny. Candidates are represented as natural numbers.  For the representation of  votes, two distinct data structures were employed to account for the differing ways in which the algorithms process voter preferences:
\begin{dafny}[numbers=none]
type Votes_Score = seq<map<nat,int>>;
type Votes_PreferenceList = seq<seq<nat>>;
\end{dafny}
In the first data structure, which is used in Score Voting, each vote is a map from candidates (natural numbers) to integers. Borda Count, IRV, and STV use the second one, where each vote is a preference list of candidates in descending order, with the highest-ranking candidates appearing first.

For the more complex algorithms, the pseudo-code was decomposed into  smaller sub-methods. This separation reduced the difficulty of reasoning about these specific algorithms by isolating intricate behavior into smaller verifiable components, a common approach in modular verification~\cite{Wimmer19}. It also facilitated collaboration within the team, as different members focused on different sub-parts of the algorithms without interfering with other colleagues.

Once the pseudo-code design was completed, the development process moved on to the implementation in Dafny. For the majority of the sub-methods, a functional model and a concrete implementation were provided and proven consistent. This approach guarantees that the executable code adheres to the higher-level specification. For larger methods that combined several sub-components, Dafny struggled with purely functional formulations. In these cases, postconditions and invariants were the key to establish correctness.

After verification was achieved at the Dafny level, Dafny’s code extraction capabilities were used to generate executable implementations in C\# and Go. This allowed the integration of the verified algorithms into an e-voting web sevice without sacrificing performance or usability, while still retaining the formal guarantees established during verification.

\subsection{Algorithm 1: Instant-Runoff Voting}
\label{sec:irv}
The verification of the Instant-Runoff Voting algorithm showcases the basics of how the verification of an election algorithm with successive elimination rounds can be accomplished.

\paragraph{Algorithm}

The result of an election is determined by applying the steps below on the candidates 
(\inl{set<nat>})
and votes (\inl{Votes_PreferenceList}, see Sect.~\ref{sec:methodology}).
The candidates are represented as 
\inl{set<nat>}
to guarantee that the order of candidates cannot influence the winner and to avoid duplicate candidates by construction.
The algorithm consists of the following steps:
\begin{enumerate}[leftmargin=*]
  \item Calculate obtained votes per candidate (how many times a candidate was given as first preference).
  \item If some candidate got an absolute majority, they are declared the winner and the algorithm terminates. Else, gather the candidates who got the lowest votes.
  \item Update candidates and votes by removing the lowest-scoring candidates.
    Evaluate the next voting round, by repeating the steps above until either a winner is found or all candidates are removed, resulting in no winner.
\end{enumerate}

\paragraph{Specification} 
The method \inl{InstantRunoffVoting}
computes the result of an election and returns either the candidate who wins or 0 to indicate no winner exists.

\begin{dafny}[firstnumber=1]
method InstantRunoffVoting(cands: set<nat>, 
 votes: Votes_PreferenceList) returns (winner: nat)
  requires forall i,j:: (0<=i<|votes| && (*@\label{irvCond1}@*)
   0<=j<|votes[i]|) ==> votes[i][j] in cands
  requires forall i,j,k:: 
   (0<=i<|votes| && 0<=k<|votes[i]| && 
   0<=j<|votes[i]| && k!=j) 
   ==> votes[i][j] != votes[i][k]
  requires 0 !in cands
  ensures |votes| == 0 ==> winner == 0 (*@\label{irvCondNoVotes}@*)
  ensures cands == {} ==> winner == 0 (*@\label{irvCondNoCands}@*)
  ensures winner != 0 ==> winner in cands (*@\label{irvCondWinIsCand}@*)
  ensures winner == 0 && |cands|!=0 ==> (*@\label{irvCondWithIswinner}@*)
   (forall c:: (c in cands) 
   ==> !isWinner_IRV(c, cands, votes))
  ensures winner !=0 ==> 
   isWinner_IRV(winner, cands, votes)
   {...}(*@\footnote{Code manually reformatted for presentation.\label{code_disclaimer}}@*)
\end{dafny}

The preconditions of the \inl{InstantRunoffVoting} method enforce that all votes are for existing candidates, no preference list contains duplicate candidates and 0 cannot be a candidate, so it can be safely returned to signal that no winner exists ~\hyperref[irvCond1]{(lines~\ref*{irvCond1}-9)}. 
The postconditions ensure that no winner can exist if either no votes or no candidates ~\hyperref[irvCondNoVotes]{(lines~\ref*{irvCondNoVotes}-11)} are given and that a returned winner is an existing candidate~\hyperref[irvCondWinIsCand]{(line~\ref*{irvCondWinIsCand})}.

The hardest part of the verification was to correctly specify the different voting rounds, as the naive approach of returning the election winner or 0 in one step proved to be  difficult. Instead, the code argues about candidates individually, using 
\inl{isWinner_IRV} \hyperref[irvCondWinIsCand]{(lines~\ref*{irvCondWithIswinner}-17)}, 
which returns true if a specific candidate wins the election.

\break
\begin{dafny}
ghost function isWinner_IRV(
 candToCheck: nat, cands: set<nat>, votes: 
 Votes_PreferenceList) : (isWinner: bool)
  ...
  ensures !isWinner && cands != {} ==> (*@\label{irvCondNoMajority}@*)
   !hasMajority_IRV(candToCheck, cands, votes)
  ensures !isWinner && cands != {} ==> (*@\label{irvCondNotWinner}@*)
  hasWinnerInCurrentRound_IRV(cands, votes)
  || candToCheck in lowestCands_IRV(cands, votes)
  || !isWinner_IRV(candToCheck, 
    cands-lowestCands_IRV(cands, votes), 
    votesWithRemovedCandidates_IRV(cands, votes,
     lowestCands_IRV(cands, votes)))
  ensures isWinner ==> |votes|!=0 &&( (*@\label{irvCondWinner}@*)
   hasMajority_IRV(candToCheck, cands, votes)||(
    candToCheck !in lowestCands_IRV(cands, votes) 
    && isWinner_IRV(candToCheck, 
     cands-lowestCands_IRV(cands, votes), 
     votesWithRemovedCandidates_IRV(cands, votes,
      lowestCands_IRV(cands, votes)))))
\end{dafny}

The most important postconditions specify this behavior by checking if a candidate got an absolute majority in the current voting round or by recursively calling \inl{isWinner_IRV} to see if they win the next voting round with updated votes and candidates. 
If a candidate does not win, they must not have an absolute majority in the current round \hyperref[irvCondNoMajority]{(lines~\ref*{irvCondNoMajority}-24)}. Additionally, either someone else wins in the current round, the candidate gets eliminated in this round, or they are not a winner in the next voting iteration \hyperref[irvCondNotWinner]{(lines~\ref*{irvCondNotWinner}-31)}. If the candidate wins, then they must have an absolute majority in the current round, or they do not get eliminated in this round and are a winner in the next voting iteration \hyperref[irvCondWinner]{(lines~\ref*{irvCondWinner}-38)}.

\paragraph{Implementation Details}
Removing lowest-voted candidates from the votes 
serves as a good example of how components of the algorithm are verified.
The desired result for a single preference list after removing the candidates is specified by \inl{singleVoteWithRemovedCands_IRV}. 
The correct removal is achieved by iterating over all preference lists, guaranteeing that the candidates were correctly removed from the lists already encountered \hyperref[irvCondOuterWhile] {(lines~\ref*{irvCondOuterWhile}-48)}.
For a single preference list, a nested while loop iterates over all candidates, only keeping the candidates not supposed to be removed \hyperref[irvCondInnerWhile]{(lines~\ref*{irvCondInnerWhile}-53)}.
\begin{dafny}
method VotesWithRemovedCands_IRV(
fullCands: set<nat>, votes: Votes_PreferenceList, 
candsToRemove: set<nat>) 
returns (newVotes:seq<seq<nat>>)
... {...
 while (i<|votes|) (*@\label{irvCondOuterWhile}@*)
 ...
 invariant forall j:: (0<=j<i) ==> newVotes[j] == 
  singleVoteWithRemovedCands_IRV(
  fullCands, votes[j], candsToRemove)
 {...
  while (j<|votes[i]|) (*@\label{irvCondInnerWhile}@*)
  invariant newVote == 
   singleVoteWithRemovedCands_IRV(
   fullCands, votes[i][..j], candsToRemove)
  {...} ...}}
\end{dafny}

\input{fig1}

An important design decision was to keep votes that no longer contain candidates as empty votes, instead of removing 
them from the votes sequence. This enables the \inl{votesWithRemovedCands_IRV} function to enforce that each preference list returned in \inl{newVotes} must be the result of \inl{singleVoteWithRemovedCands_IRV} of the corresponding vote from the input. If empty votes got removed instead, it is hard to specify the expected result, because either the vote could 
become empty if all listed candidates should be removed, or the index 
of the corresponding new vote is different from its starting index. 
One minor disadvantage introduced by this decision is 
that the required vote count for an absolute majority can no longer use the total number of votes as a baseline. It rather requires the
amount of non-empty votes; however, this can easily be computed.

\subsection{Algorithm 2: Single Transferable Vote}

\contents{
\item A second algorithm in more detail
\item STV?
}
\paragraph{Algorithm}
The Single Transferable Vote system extends the idea of Instant-Runoff Voting to multi-seat elections. 
It ensures proportional representation by redistributing surplus votes from already elected candidates and eliminating the lowest-scoring candidates until all seats are filled.

The algorithm operates on a list of candidates 
(\inl!seq<nat>!)
and a list of votes 
(\inl{Votes_PreferenceList}, see Sect.~\ref{sec:methodology})
without empty ballots.
Candidates are represented as a sequence to enable deterministic tie-breaking: if multiple candidates obtain the same number of votes, the candidate appearing earlier in the sequence is selected for elimination or surplus redistribution, ensuring deterministic behavior. The algorithm computes the subset of candidates that will occupy the specified number of seats. In this way, every candidate who was assigned a seat represents a winner. Furthermore, the method employs the AUTOFILL procedure, which assigns a seat to every remaining candidate and classifies them as winners if the number of remaining candidates equals the number of unassigned seats.
The main steps are follows:
\begin{enumerate}
  \item Count first-preference votes for each candidate.
  \item If the number of remaining candidates equals the number of unfilled seats, assign one seat to each of them.
  \item Otherwise, determine the candidate with the highest vote count.
  \item If this candidate meets or exceeds the Droop quota 
  $(\frac{|Votes|}{seats + 1} + 1)$, classify them as elected, redistribute their surplus votes proportionally and remove them from the list of votes and candidates.
  \item If no candidate reaches the quota, eliminate the lowest-scoring candidate, redistribute their votes and remove them from the list of votes and candidates.
  \item If at least one seat remains unassigned, return to step~2. Otherwise, the winners have been determined.
\end{enumerate}

\paragraph{Specification}
The method 
\inl{SingleTransferableVote}, shown in Fig.~\ref{fig:transferMethods},
computes the result of a multi-seat election and returns the elected candidates. The preconditions of the \inl{SingleTransferableVote} method enforce that all votes are non-empty, contain only registered candidates, and have no duplicates. Additionally, the candidate list must not be empty or contain duplicates, and the number of seats to be filled must not exceed the number of candidates.

Since the algorithm includes the AUTOFILL procedure, the wrapper method distinguishes between candidates elected through quota classification and those selected by AUTO\-FILL. This separation preserves the order of quota winners according to the time of their classification, which enables the verification of further properties.

Under these conditions, the postconditions ensure that all winners originate from the list of registered candidates, and that the total number of winners---both quota winners 
(\inl{W}) 
and AUTOFILL winners 
(\inl{Rest})---equals the available number of seats ~\hyperref[stvSum]{(line~\ref*{stvSum})}. Furthermore, the sequences 
\inl{W} 
and 
\inl{Rest}
are proven to be disjoint, and each candidate in 
\inl{W} 
has at least one supporting vote ~\hyperref[stvVote]{(line~\ref*{stvVote})} and reaches the quota at classification ~\hyperref[stvScore]{(line~\ref*{stvScore})}. The postcondition used to verify this property states that every winner \inl{c} contained in \inl{W} has an associated vote list, candidate list, and factor list, ensuring that its score computed from these is at least the quota.

To support this verification, the wrapper method records each classification step in 
\inl{stateSet}, which is
a sequence of triples containing the current vote list, factor list, and candidate list at the time of classification. 
Thus, every candidate 
\inl{c} 
at a valid index 
\inl{i} 
in 
\inl{W} 
has a corresponding state in 
\inl{stateSet} 
at the same index. This variable is used solely for verification and is omitted from the compiled code; it can be seen as a witness for the correct execution of the algorithm.

The method \inl{TransferVotes}, shown in Fig.~\ref{fig:transferMethods}, 
ensures correct proportional redistribution of votes. Here, \inl{F} 
represents the factor list, 
\inl{q} 
the quota, and 
\inl{m} 
the total votes of candidate 
\inl{c}. 
If a candidate has not reached the quota, the factor of every transferable vote with the given candidate as first preference is carried over unchanged into the output factor list ~\hyperref[stvTV1]{(line~\ref*{stvTV1})}; otherwise, the factor of every transferable vote with the given candidate as first preference is adjusted by the ratio of surplus to total votes and placed in the new factor list ~\hyperref[stvTV2]{(line~\ref*{stvTV2})}. In this postcondition, the difference \inl{m-q} represents the number of surplus votes, whereas the function \inl{getIndicesSet(...)} computes a set of votes containing the candidate \inl{c} as first preference, and the cardinality of this set corresponds to the total number of votes achieved by \inl{c}. Votes that become empty are treated as non-transferable.

\paragraph{Implementation Details}
The algorithm is implemented in a modular manner, with sub-methods for candidate elimination, vote evaluation, candidate search, vote redistribution, and a wrapper method implementing STV by invoking the necessary sub-methods. 
Important design decisions were to separate quota winners from AUTOFILL winners, enabling verification of the stated correctness properties, and to record all classification steps in order to capture each candidate's score at the time of classification. This is outlined in Fig.~\ref{fig:transferMethods}.

\section{Deployment as Web Application}

\contents{
\item Architecture
\item Enforcement of Preconditions
\item Handling of invalid Elections
}
Multiple approaches and technologies were used for the implementation of
a service that uses verified Dafny code at its core to hold elections online.
Dafny was transpiled into the programming language Go to be used in a gRPC server~\cite{Go-grpc}\footnote{Currently hosted on \url{https://eldarica.org/eins/}} and into C\# for use in a Blazor Server app~\cite{blazor-doc}.\footnote{Currently hosted on \url{https://eldarica.org/superior/}} Since Blazor is already well documented by Microsoft, we will focus on the architecture behind our gRPC server.
\begin{figure}[h] 
    \centering
    \includegraphics[width=\linewidth]{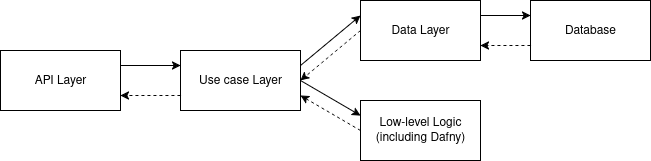} 
    \caption{The architecture of the Go backend}
    \label{fig:archGo}
\end{figure}

\paragraph{Architecture}
The architecture (see Fig.~\ref{fig:archGo}) of the gRPC Go backend is similar to an onion architecture. The dependencies only flow outward; e.g., the domain logic (Use case Layer) can use the data loaders, but the database cannot depend on the fact that gRPC is used instead of a REST API. The different layers are as follows (with the outermost first):
\begin{itemize}
    \item API Layer: Implements the gRPC functions defined by a Protobuf.
    \item Use case Layer: Contains domain logic in a broad sense. Also glues together low-level logic calls and data calls.
    \item Data Layer: Provides functions to interact with the database as well as data objects.
    \item Low-level Logic Layer: Contains independent low-level functions, such as password hashing. The formally verified voting algorithms with their caller functions are also included in this layer.
\end{itemize}

\paragraph{Enforcement of Preconditions}
For the enforcement of preconditions, two distinct approaches have been used in the different backends.

The first approach is that preconditions are checked before the transpiled Dafny code is called. This allows the code to be optimized by hand, and if a check fails, an appropriate error message and behavior can be triggered. However, the main disadvantage is that the enforcing code needs to be checked and tested rigorously, because if there is a bug, the Dafny code might not perform as expected. This approach was used only in the Go implementation.

The second approach is that Dafny is solely responsible for enforcement at runtime via the \inl{expect} keyword.
If any precondition in Dafny does not hold, the transpiled code throws a Dafny.HaltException in C\# or a runtime panic in Go, which is then caught. The exception message is customizable and thus increases the informational value, for example in server log files. A main disadvantage is that the Dafny library is a black box for the server, which must blindly trust that it executes correctly.
It is also important that the runtime check (\inl{expect}) and the formal preconditions (\inl{requires}) are rigorously checked to ensure that they are consistent.
This approach was used in both implementations.

\paragraph{Handling of Invalid Elections}
If a post- or precondition does not hold, the server either stops the evaluation before the transpiled Dafny code is called
or catches an exception or panic afterwards. The election is then deleted from the database and an e-mail is sent to all registered voters for complete transparency.

\section{Lessons Learned}

\contents{
\item Sometimes painful Dafny learning experience
\item Scalability issues with large programs and complicated specs
}

\paragraph{Focus on Simplicity}
Keeping implementation and specification as simple as possible was a huge help in many cases, for instance when decoding the votes 
for the scoring voting systems where each candidate can be assigned a score. Initially, this was done as a sequence of tuples, 
which contained the candidate and their score. However, using a map from candidates to their scores simplified preconditions, because it was no longer required to check if a candidate appears multiple times in a vote. Furthermore, being able to better access the score 
of a candidate in a vote helped in the formulation of related pre- and postconditions and saved considerable work elsewhere, as one fewer precondition needed to be proven in order to call different methods.
Another example is explained in Sect.~\ref{sec:irv}, choosing to keep empty votes as empty lists when removing candidates from votes.

\paragraph{Performance and Scalability}
Reasoning about properties of voting algorithms frequently required reasoning about entire collections, such as all candidates or all votes. Consequently, method contracts and invariants relied heavily on quantified statements, slowing the interactive verification process down and leading to numerous solver timeouts. To overcome this hurdle, we introduced a temporary \inl{assume false} statement to incrementally verify methods. This statement allows the solver to focus only on the logic preceding the statement by causing all logic following it to be trivially true. Our process now involved progressively moving this statement downwards through the method body until the entire method was proven and the \inl{assume} could be removed. Furthermore, we guided the solver through difficult proofs by introducing strategic \inl{assert} statements to speed up verification.

Separately, we identified a significant performance issue related to Dafny's process management. To check satisfiability of a program, Dafny runs Z3 Theorem Prover processes in the background. We observed that after a verification run finished, according to Dafny, some of these processes would not always terminate, especially after rapid code changes. While these orphaned processes would not influence future verification runs, their accumulation caused excessive memory consumption and high CPU loads, impacting the overall system performance heavily. 
We found manually terminating the rogue Z3 processes through the operating system's task manager after a few verification runs to be the most effective solution.

\paragraph{Automation and Manual Intervention}
Some properties that are intuitively obvious for humans required manual guidance for the underlying solver via specific \inl{assert} statements or detailed \inl{lemma}.
Pinpointing the exact ``obvious'' fact the solver was missing turned out to be an unexpected challenge by itself.
The following code snippet illustrates this. Here, $S$ is a sequence of some kind:
\begin{dafny}[numbers=none]
var A:= S[..|S|];
assert S == S[..|S|]; //necessary
assert multiset(S) == multiset(A);
\end{dafny}
For a human, it is obvious that $A$ and $S$ are identical and share the same \inl{ multiset}, but the solver cannot prove the final assertion without adding the intermediate one.

\section{Conclusions and Future Work}
In this paper, we have presented a library of four prominent voting algorithms and verified their key postconditions within the Dafny framework, focusing in detail on Instant-Runoff Voting and Single Transferable Vote. These verified algorithms serve as the trusted  core for our deployed e-voting websites. The  code base has been made publicly available~\cite{Verification_of_E-Voting_2025}.

Immediate next steps for further development of our work are to verify more postconditions and fairness criteria for the existing algorithms using our code as a baseline.
A more advanced goal is the expansion of the collection of verified algorithms using our approach. This could include a different  social choice function like a Condorcet method~\cite{Fishburn77}.

\subsubsection*{Acknowledgments}

We want to thank the anonymous reviewers for helpful feedback. This work was supported by the COST action CA20111 EuroProofNet and by the Swedish Research Council (VR)
    under grant~2021-06327.

\bibliographystyle{ACM-Reference-Format}
\bibliography{papers}

\end{document}

%% file: fig1.tex
\begin{figure*}[t]
\begin{dafny}
type Triple = (Votes_PreferenceList,seq<real>,seq<nat>)
predicate ValidInputsSTV(V : Votes_PreferenceList, C : seq<nat>){  (*@\label{stvPreCond}@*)
 (0 < |C| && (forall c :: c in C ==> multiset(C)[c] == 1) && (forall i :: 0 <= i < |V| ==> V[i] != [])
 && (forall v,c :: v in V && c in v ==> (c in C && multiset(v)[c] == 1))}
 
method SingleTransferableVote(V : Votes_PreferenceList, C : seq<nat>, s: nat)
returns (W : seq<nat>, Rest : seq<nat>, stateSet : seq<Triple>)
  requires s <= |C|
  requires ValidInputsSTV(V, C)
  ensures |W| + |Rest| == s (*@\label{stvSum}@*)
  ensures forall c :: c in W ==> exists v :: v in V && c in v (*@\label{stvVote}@*)
  ensures forall i :: 0 <= i < |W| ==> (*@\label{stvScore}@*)
    calculateTotalValue(stateSet[i].0, stateSet[i].2, stateSet[i].1, W[i])
      >= ((((|V| as real)/(s + 1) as real)) + 1.0).Floor as real 
{
  var q := (((((|V| as real)/ (s + 1) as real)) + 1.0).Floor) as real;
  W, Rest, stateSet := [], [], [];
  var F : FactorList := seq(|V|, i => 1.0);
  var F0, C0, V0 := F, C, V;
  while |W| + |Rest| < s {
    if |C0| == s-|W| { Rest := C0; } // Step 2
    else {
      // Create ranking R with first-preference votes for candidate (step 1), get maximum score from R
      m := max(R.Values);
      // Candidate did not achieve quota --> find the loser (step 5)
      if m < q { m := min(R.Values); }
      // Find the candidate c with score m
      if m >= q {
        W := W + [c]; stateSet := stateSet + [(V0, F0, C0)]; // classification in step 3
      }
      // Vote redistribution and elimination from step 4 and 5: remove the candidate c from C0, V0 and
      // invoke TransferVotes()
}}}

method TransferVotes(V : seq<seq<nat>>, F : seq<real>, c: nat, q : real, m : real)
returns (A0: (Votes_PreferenceList, FactorList))
  requires |V| == |F|
  requires m > 0.0 ==> exists i :: 0 <= i < |V| && V[i][0] == c
  ensures forall i :: 0 <= i < |V| && V[i][0] == c && |V[i]| > 1 && !(m - q >= 0.0) ==> F[i] in A0.1 (*@\label{stvTV1}@*)
  ensures forall i :: 0 <= i < |V| && V[i][0] == c && |V[i]| > 1 && m - q >= 0.0 ==> (*@\label{stvTV2}@*)
    F[i]*((m-q)/|getIndicesSet(V,c)| as real) in A0.1 
  {...}
\end{dafny}
\caption{Methods \inl{SingleTransferableVote} and \inl{TransferVotes}}
\label{fig:transferMethods}
\end{figure*}